\documentclass[final,5p,times,twocolumn]{elsarticle}

 \usepackage{graphicx}
\usepackage{amssymb}
\usepackage{epstopdf}
\usepackage{caption}
\usepackage{subcaption}
\usepackage{longtable}
\usepackage{rotating}
\usepackage{lscape}
 \usepackage{lineno}
  \usepackage{epsfig}
  \usepackage{subfig}
  \usepackage{url}
 \biboptions{comma,square}

\journal{Journal of Environmental Protection}

\begin{document}

\begin{frontmatter}



\title{Measurements of Fission Products from the Fukushima Daiichi Incident in  San Francisco Bay Area  Air Filters, Automobile Filters, Rainwater, and Food}


\author[LBNL]{A.R. Smith}
\author[LBNL,UCB]{K.J. Thomas\corref{cor1}}
\author[LBNL,UCB]{E.B. Norman}
\author[LBNL]{D.L. Hurley}
\author[UCB]{B.T. Lo}
\author[LBNL]{Y.D. Chan}
\author[USP]{P.V. Guillaumon}
\author[LBNL]{B.G. Harvey}

\address[LBNL]{Nuclear Science Division, Lawrence Berkeley National Laboratory, Berkeley, CA 94720}
\address[UCB]{Department of Nuclear Engineering, University of California, Berkeley, CA 94720}
\address[USP]{Department of Physics, University of S\~{a}o Paulo, S\~{a}o Paulo- SP 05314970 , Brazil}
\cortext[cor1]{Corresponding Author. Email: kjthomas@lbl.gov}

 \linenumbers
 
\begin{abstract}
A variety of environmental media were analyzed for fallout radionuclides resulting from the Fukushima nuclear accident by the Low Background Facility (LBF) at the Lawrence Berkeley National Laboratory (LBNL) in Berkeley, CA. Monitoring activities in air and rainwater began soon after the onset of the March 11, 2011 tsunami and are reported here through the end of 2012.  Observed fallout isotopes include  $^{131}$I, $^{132}$I,$^{132}$Te,$^{134}$Cs, $^{136}$Cs, and $^{137}$Cs. Isotopes were measured on environmental air filters, automobile filters, and in rainwater. An additional analysis of rainwater in search of $^{90}$Sr is also presented. Last, a series of food measurements conducted in September of 2013 are included due to extended media concerns of $^{134, 137}$Cs in fish. Similar measurements of fallout from the Chernobyl disaster at LBNL, previously unpublished publicly, are also presented here as a comparison with the Fukushima incident. All measurements presented also include natural radionuclides found in the environment to provide a basis for comparison.
\end{abstract}

\begin{keyword}
Fukushima Daiichi Nuclear Power Plant, Fukushima, fallout, air monitoring, rainwater, automobile filters, Chernobyl
\end{keyword}

\end{frontmatter}


%\linenumbers

\section{Introduction}
Since the early 1980's, the Low Background Facility (LBF) at Lawrence Berkeley National Laboratory (LBNL) has been analyzing laboratory environmental air sampler filters for the presence of any  gamma-emitters, such as naturally occurring, $^{7}$Be and $^{210}$Pb.  After the announcement of the accident at the Fukushima Daiichi Nuclear Power Plant, detailed monitoring began to search for fallout isotopes in environmental air filters at the local LBF station at LBNL and in rainwater in Oroville, CA. Several other groups have reported similar Fukushima-related measurements in air filters and rainwater in the U.S. \citep{Norman2011, Bandstra2011, Leon2011, MacMullin2012, Biegalski2012}. The fallout also provided a demonstration for a pilot program in monitoring fallout isotopes in automobile filters, which are also continuously monitored by the LBF. An additional analysis was later performed upon rainwater collected in the spring of 2011 in an effort to search for the presence of $^{90}$Sr in the San Francisco Bay Area. Authors involved in this study, A.R. Smith and E.B Norman, also performed similar measurements in California in the aftermath of the Chernobyl accident in 1986 which were never published formally, and are presented here in comparison to those on Fukushima.  Last, due to extended reports and local concerns of contaminated water being leaked in 2013 into the ocean at Fukushima, a series of food measurements were performed in September of 2013 to search for fission products in fish and other food items from the Pacific region purchased at local Bay Area retail locations.

\subsection{The Low Background Facility at Lawrence Berkeley National Laboratory}
The LBF at LBNL typically performs  a variety of low background gamma spectroscopy services to a variety of experiments and end users. The LBF operates in two unique, low-background laboratory spaces: a local surface laboratory constructed of low-activity concrete; and an underground site with over 500 m.w.e (meters water-equivalent) overburden for shielding against cosmic ray muons.  A primary component of its activities include low background counting of candidate construction materials for ultralow background experiments such as those searching for dark matter, neutrinos, and neutrinoless double beta decay -- which often have strict requirements for radiopurity. Of primary concern in these experiments are the natural primordial radioisotopes and their decay chains ($^{238,235}$U,$^{232}$Th, $^{40}$K), common man-added radioactivity ($^{60}$Co, $^{137}$Cs, etc.), and various cosmogenic radioisotopes. Other services include neutron activation analysis for trace analysis of many stable isotopes, neutron flux/transmutation doping, radon emanation, R\&D hosting, environmental measurements, and other various applications such as accelerator characterization and radiological waste assay. More information on the Low Background Facility at LBNL can be found in \citep{Thomas2013} and \citep{ThomasLRT}.

\section{Air Sampling}
An air sampler was put into operation outside LBNL LBF on 3/14/2011, to track the arrival and continued presence at LBNL of gamma-emitting fission products from the Fukushima Daiichi nuclear power plant disaster that was caused by the earthquake/tsunami on 3/11/2011.  This sampling station has continued operation to the present and results through the end of 2012 are summarized here. 

\begin{figure}[!ht]
\centering
\includegraphics[width=\linewidth, angle=0]{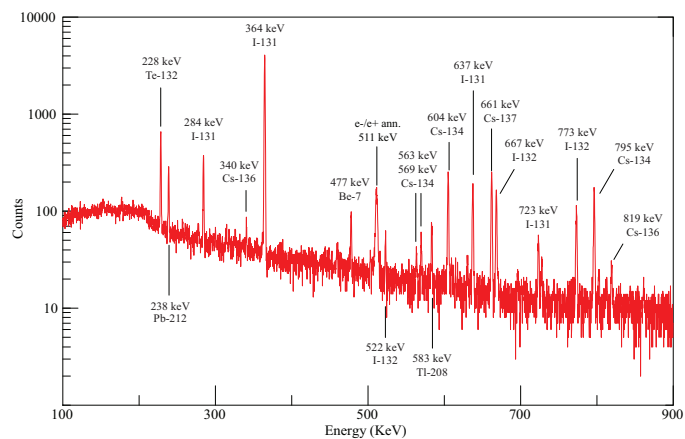}
\caption{A HPGe gamma ray spectrum from an air filter exposed March 23-24, 2011 at the LBF facility at LBNL containing both natural and fission nuclides. This spectrum represents a full 24 hour exposure period and was counted for a livetime of 1250 minutes immediately upon being removed from the air sampler.  More information on the peaks displayed here are listed in {\bf Table \ref{tab:isotopes}}.}
\label{fig:airfilter}
\end{figure}

 Air sampling began on the morning of 3/14/11, using a nominal 24-hr collection period, followed by immediate counting on an ORTEC low-background packaged HPGe detector (2.26 kg n-type HPGe crystal,  relative efficiency of 115\%). Air flow through the sampler is actively regulated at a constant flow rate, producing 244.7 m$^{3}$ throughput in a 24 hour collection period. Aerosols are collected upon a 10.16 cm diameter HEPA filter rated for 0.3 $\mu$m aerosol sizes and counted directly atop the top face of the detector. For the sake of these measurements, it was assumed that essentially all fission products in the air are attached to aerosols greater than this size. All filters analyzed with the n-type HPGe detector showed statistically significant peaks from the naturally produced airborne gamma-emitters, such as $^{7}$Be and $^{210}$Pb.  Initial counting (within an hour after end of sampling) of filters also showed prominent peaks from the short-lived daughters of U-series $^{222}$Rn  and Th-series $^{220}$Rn.  Analysis of gamma ray peaks from  short-lived Rn daughters could provide useful information with respect to collection of fission product nuclides, especially for short collection periods. The analysis procedure included several short counting intervals to quantify Rn daughter activities, followed by an overnight count to increase the possibility of detecting very small peaks, as may be expected from the first signs of fission product radionuclides from the damaged Japanese reactors.  Filter exchanges were initially performed in 24 hour cycles. Sampling intervals were later extended to 2 or more days, one week, and eventually to one month in order to continue to produce actual values for the decreasing activities of gamma-emitters rather than only upper limits.  Through this strategy, activities over a range of 1000-fold per 24 hour collection period were documented for gamma-emitters from which the maximum observed count rate in diagnostic peaks were as low as 1 count per minute. Throughout the monitoring period, observed fallout isotopes include $^{131}$I (t$_{1/2}$ = 8.0 days), $^{132}$I (t$_{1/2}$ = 2.3 hours) ,$^{132}$Te (t$_{1/2}$ = 3.2 days),$^{134}$Cs (t$_{1/2}$ = 2.06 years), $^{136}$Cs (t$_{1/2}$ = 13.16 days) and $^{137}$Cs (t$_{1/2}$ = 30.07 years). A sample gamma spectrum from such an air filter is displayed in {\bf Figure \ref{fig:airfilter}}. 

\begin{figure}[!ht]
\centering
\includegraphics[width=\linewidth, angle=0]{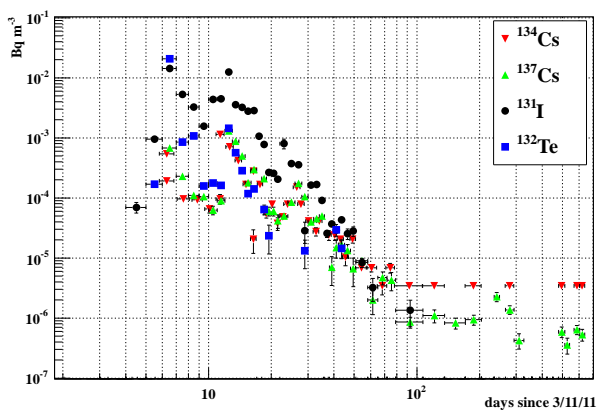}
\caption{Long term airborne concentrations fission products from March 11, 2011 to the end of 2012 as measured on HEPA filters at the LBF in Berkeley, CA. Horizontal error bars, when visible, represent filter exposure periods. Missing error bars are smaller than the visible data marker. Plotted uncertainties seen here are statistical only, and a conservative systematic uncertainty of $\sim$10\% should also be assumed.}
\label{fig:allfalloutfull}
\end{figure}

The overnight count of the filter collected on March 15-16, 2011 showed the initial presence of a very small peak at 364 keV, the most intense gamma emitted in the decay of $^{131}$I, but without any companion fission products could not provide unequivocal evidence to support official arrival. When alone, the $^{131}$I could have originated from a nearby source, such as a local hospital. The following day provided the full confirmation of fission products from Fukushima from a filter exposed March 16-17, in which $^{132}$Te and $^{131}$I were both observed. The filter collected during March 17-18, 2011 showed the highest activities observed for $^{131}$I and $^{132}$Te, at 14.3$\pm$0.1 mBq m$^{-3}$ and 20.9$\pm$0.1 mBq m$^{-3}$ respectively, as well as the first peak in their temporal distributions. The filter collected from March 23-24 exhibited the second highest $^{131}$I and $^{132}$Te activities, at 12.5$\pm$0.1 mBq m$^{-3}$ and 1.44$\pm$0.03 mBq m$^{-3}$ respectively, and may represent the Òsecond lapÓ of the most radioactive plume as it passed by California again.  Note all uncertainties listed here are statistical only, and a $\sim$10\% systematic uncertainty should be assumed as well. The relative temporal and absolute airborne activities based upon the air filter measurements are shown for the full duration of the LBF's Fukushima fallout monitoring activities through the end of 2012 in {\bf Figure \ref{fig:allfalloutfull}}. Not plotted are $^{136}$Cs, since it only was present on a small amount of filters, and $^{132}$I, since it is the short-lived daughter of $^{132}$Te.  The activities of $^{210}$Pb and $^{7}$Be, both naturally occurring radioisotopes found in the air, are shown in {\bf Figure \ref{fig:pbbe}}. The natural isotopes can be used as a comparison to the Fukushima fallout in {\bf Figure \ref{fig:allfalloutfull}} to demonstrate that the observed fission radioisotope activities at the LBF station were, briefly at their initial arrival, comparable to natural radioisotopes that are always present, but fell much below natural background activities quickly. In addition, it is also worth noting that only $^{210}$Pb and $^{7}$Be were plotted, and there are many other $^{222}$Rn daughters such as $^{214}$Bi not shown in {\bf Figure \ref{fig:pbbe}} since most $^{222}$Rn daughters are quite short lived. The measurements of airborne fallout activities presented here are also in reasonable agreement with the results of another local group at UC Berkeley who made measurements in the area \citep{Bandstra2011}.

%\begin{figure}[!ht]
%\centering
%\includegraphics[width=\linewidth, angle=0]{AllFalloutFullloglin-first30days}
%\caption{The first 35 days of airborne fallout monitoring at the LBF as shown in days since March 11, 2011 as measured on HEPA filters at the LBF in Berkeley, CA. The arrival at the LBNL LBF began with $^{131}$I on March 14. Horizontal error bars, when visible, represent filter exposure periods. Missing error bars are smaller than the visible data marker. Plotted uncertainties seen here are statistical only, and a conservative systematic uncertainty of $\sim$10\% should also be assumed.}
%\label{fig:allfallout30}
%\end{figure}

\subsection{Automobile Air Filters}
Since 2002, the Low Background Facility has been analyzing automobile filters obtained from the local Berkeley Police Department to perform regular analysis as a pilot program for detecting radioisotopes with potential homeland security applications. Over 1200 automobile filters have been counted over the course of the program. The release of fission products from Fukushima have allowed for a `proof-of-principle' for this radioisotope monitoring technique. Although they do not have the collection efficiency of HEPA filters (automobile filters collect at approximately one-third the efficiency as the 0.3 $\mu$m filters used in the previous section), they can still produce reliable airborne measurements since their airflow can even be estimated based upon the odometer reading and fuel consumption of a vehicle. However, the use of automobile filters perhaps provides more important qualitative advantages. First, with the use of public sector vehicles that run somewhat regular patrol routes, there is a low-cost network of filters \emph{already deployed} and screening essentially every city. With regular screening of filters removed from vehicles at normal maintenance exchanges, any unusual radioactivity found in an air filter could trigger `first-alert' procedures that would call for further investigation. After determining the true presence of an isotope of concern, routes from logbooks can be determined and teams could be scrambled somewhat quickly to perform more thorough sweeps of areas looking for potential sources of radioactivity simply by driving vehicles and quickly analyzing filters upon return, or perhaps with more sophisticated portable filtration and detection systems. Second, in the event of a large-scale disaster, networks of automobiles could be deployed to determine the level and extent of contamination across large regions and could be analyzed via a large network of existing detectors at laboratories and universities nationally.

The Low Background Facility at LBNL obtains automobile filters from the Berkeley Police Department every few weeks. At regular maintenance intervals, air filters are removed and (ideally) the odometer readings are recorded. In the LBF pilot program, the filters are initially screened at the vehicle maintenance shop using a First Response Detector (FRD) composed of a 2x2 NaI detector within a 2 inch thick steel shield. Here, only gross counts of filters are noted. If any filters were to exhibit gross counting rates above background, such filters would take priority in screening. In this study no filters have ever registered above the normal variation of background readings on the FRD-- even those with radioisotopes from Fukushima. Once returned to the LBF, the filters are counted on an HPGe system in which isotopes are identified and quantified. In a full-scale operation, one could imagine placing many FRD systems at maintenance shops around the country, in which mechanics could test the filter and could flag unusual filters for additional laboratory study.

\begin{figure}[!ht]
\centering
\includegraphics[width=\linewidth, angle=0]{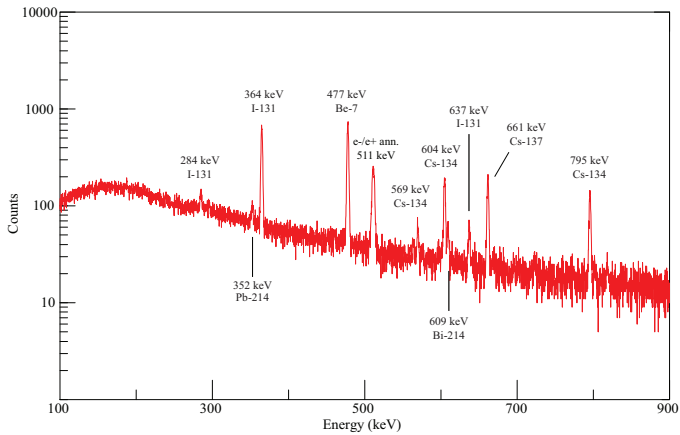}
\caption{A gamma ray spectrum from April 7, 2011 obtained from counting an automobile filter from a Berkeley Police Department patrol vehicle. Fission products were easily identifiable in the aftermath of the radiation release. More information on the peaks displayed here are listed in {\bf Table \ref{tab:isotopes}}.}
\label{fig:autofilter}
\end{figure}

\begin{figure}[!ht]
\centering
\includegraphics[width=\linewidth, angle=0]{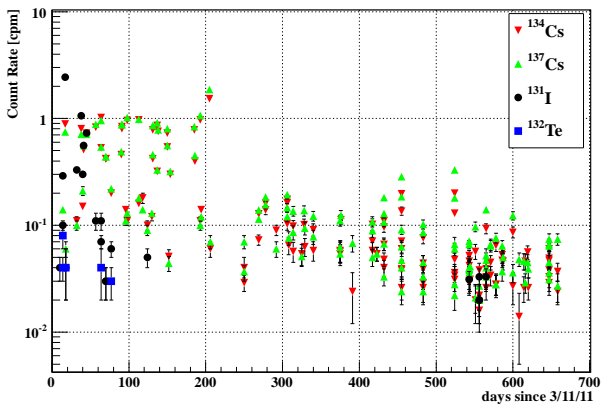}
\caption{Counting rates for fission products from Fukushima as seen on automobile filters in Berkeley, CA. The day plotted for filters represents the date after 3/11/11 when the filter was removed from a vehicle. Counting rates were not converted to airborne activity, as the full mileage (and hence the airflow) information was not available for the full data set, so the information is presented as a qualitative demonstration of the use of automobile filters for radionuclide monitoring.}
\label{fig:auto}
\end{figure}

Along with a similar program operating in Butte County at the Oroville location of the LBNL LBF, more than 1500 auto filters have been evaluated. Prior to the Fukushima disaster, the only man-added radionuclide ever observed on any filters was the rare occurrence of very small amounts of $^{137}$Cs, a relic of atmospheric nuclear weapons testing in the mid 20th century, via re-suspension of surface soil particles. No filter has ever triggered above the background threshold upon the FRD detector, even with the added Fukushima fallout. Nevertheless, the Fukushima Daiichi release of fission products provided a satisfactory test and proof-of-principle for using automobile filters as an early detection system. Although the fallout in the San Francisco Bay Area was comparable to other natural background radioactivity found in the air (such as the $^{214}$Bi, $^{210}$Pb, $^{7}$Be, etc.), several gamma ray peaks were very easily identifiable on auto filters as seen in {\bf Figure \ref{fig:autofilter}}. A plot showing the counting rates of $^{134}$Cs, $^{137}$Cs, $^{131}$I, and $^{132}$Te from the onset of the radiation release in March 2011 through December 2012 is shown in {\bf Figure \ref{fig:auto}}, with corresponding counting rates from natural radioisotopes provided in {\bf Figure \ref{fig:autonatural}}. The counting rates for automobile filters can be converted to true airborne activities using geometrical efficiency calibrations, airflow data, and collection efficiency. However, the data presented here is intended as merely a qualitative demonstration, since mileage information for the full set of filters was unavailable. By comparison to the LBF HEPA filter data in {\bf Figure  \ref{fig:allfalloutfull}}, a rough idea of the sensitivity can be established. After approximately the 100 day mark from 3/11/2011, the $^{134}$Cs and $^{137}$Cs activities in {\bf Figure \ref{fig:allfalloutfull}} from HEPA filters dropped below 10$\mu$Bq m$^{-3}$ and as seen in {\bf Figure \ref{fig:auto}}, the same isotopes still had quantifiable counting rates above background in the automobile filters. A second demonstration of detection appeared in a set of filters removed in early September of 2012 in {\bf Figure \ref{fig:auto}} where $^{131}$I reappeared, likely released as a result of a local medical procedure. This demonstrates the ability of this pilot program to detect even small amounts of radioactivity. A stationary monitor, say in the center of a city on a tall building, may not see such a release as a result of dilution and distance from the source. However, in this scenario, a police cruiser was likely near the hospital in which this isotope was released as part of a medical treatment. However, it should be noted, short lived isotopes including $^{131}$I (t$_{1/2}=$ 8 days) could be difficult to detect using automobile filters, since they can decay away before regular maintenance may occur to replace filters in vehicles.

\section{Rainwater Measurements}
\begin{figure}[!ht]
\centering
\includegraphics[width=\linewidth, angle=0]{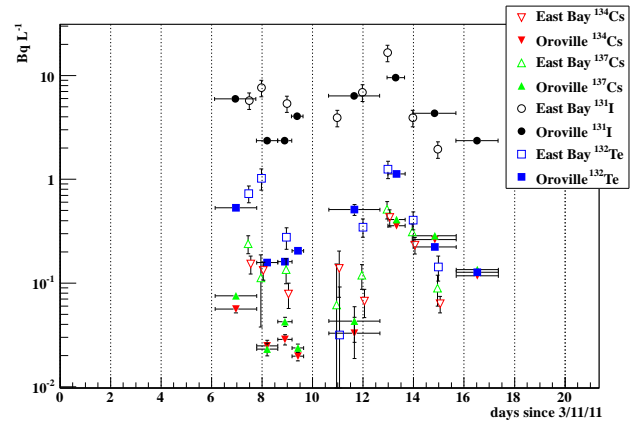}
\caption{Comparison of rainwater samples collected during the same time period at two locations approximately 120 miles apart. The hollow points represent rainwater collected in the east San Francisco Bay Area by Norman et. al. and replotted from \citep{Norman2011}. The solid points represent rainwater collected and analyzed by the LBF in Oroville, CA as seen in {\bf Figures \ref{fig:raincompare}} and {\bf \ref{fig:ororainloglog}}. The data shows remarkably similar activities in both locations, despite their distance form one another and time between rainstorms passing through each.}
\label{fig:raincompare}
\end{figure}

Also collected in the aftermath of the Fukushima announcement was rainwater in Oroville, CA. Collection periods were recorded and samples were transferred from an outdoor collector to a Marinelli-style counting beaker for counting. No other preparation was performed on the samples prior to being counted on an ORTEC HPGe spectrometer (2.1 kg p-type HPGe crystal, 85\% relative efficiency) at the underground location of the Low Background Facility with over 500 m.w.e. overburden. When applicable, results were decay corrected for the shorter lived isotopes that for decay prior to counting in order to report the activity at the time of collection. The  rainfall measurements collected in the immediate days following the Fukushima nuclear accident are shown in {\bf Figure \ref{fig:raincompare}} and the full period of monitoring in {\bf Figure \ref{fig:ororainloglog}}. The activities found in the rainwater in Oroville, CA were remarkably close in activity to the measurements made by Norman, et. al. \citep{Norman2011} on rainwater collected in the San Francisco Bay Area, which is around 120 miles away-- which is quite interesting, and are included in {\bf Figure \ref{fig:raincompare}} for comparison. This suggests that the radioactivity was spread uniformly in the precipitation as the storm front passed through northern California over the course of several hours.

\section{The Search for Strontium-90}
The rainwater samples in Norman, et al. \citep{Norman2011} contained the man-made isotopes $^{131}$I, $^{132}$I,$^{132}$Te,$^{134}$Cs,$^{136}$Cs, and $^{137}$Cs. They were collected between March 16-26, 2011 in Oakland, Albany, and Berkeley, California on the eastern side of the San Francisco Bay. Samples were placed directly into Marinelli beakers for gamma ray counting, with no additional preparation prior to gamma ray counting on a 60\% relative efficient HPGe gamma spectrometer. An efficiency calibration utilizing natural La, Lu, and K isotopes was used for analysis \citep{Perillo1996}. The samples were left sealed and archived in the laboratory after analysis, where they remained until the summer of 2012.  The highest activity found in these samples was that of $^{131}$I which peaked at 16 Bq/L on 3/24/2011. This period of rainfall also exhibited the highest activities of the other detected isotopes. 

These rainwater samples presented in Norman, et al. \citep{Norman2011} were reexamined during the summer of 2012 for the presence of $^{90}$Sr. Isotopes of strontium, in particular, are a health hazard as they are readily absorbed into bones due to its chemical similarity to calcium. Most of the fallout from Fukushima is easily observed due to the presence of high energy gamma rays which make detection via HPGe gamma spectroscopy trivial for most systems, especially those optimized for low background counting. The fission product $^{90}$Sr (t$_{1/2}$ = 28.8 years) however is much more difficult to detect. It decays via $\beta$- with Q$_{\beta}$ = 546 keV -- and \emph{no subsequent emission of gammas}. The low Q$_{\beta}$ and no gamma ray emission makes for a much trickier isotope to detect.  However, the daughter, $^{90}$Y (t$_{1/2}$ = 2.67 days),  also decays via $\beta$- with very weak emission of gamma rays, but has a much higher end point energy of Q$_{\beta}$ = 2.280 MeV which makes detection easier.

The method used to isolate $^{90}$Sr involved introducing stable strontium to the rainwater samples and creating a precipitate, which would act as a carrier to also extract any $^{90}$Sr that may be present. Four rainwater samples, each approximately 1L in volume, collected between March 18-23 in Albany, Oakland, and Berkeley CA were selected for $^{90}$Sr testing. A control sample of tap water was also analyzed. Approximately 100-140 mg of SrCl$_{2}$ was dissolved in each sample followed by 200-250 mg  of K$_{2}$CO$_{3}$ , which immediately turns the water sample a cloudy white color due to the formation of a SrCO$_{3}$ precipitate. The precipitate was then filtered out of solution onto a paper filter and dried, which was weighed before and after to obtain a precipitate mass. For every 1 mg of SrCl$_{2}$, it is expected to produce 0.931 mg of SrCO$_{3}$. The experimental conversion we achieved averaged 91.61\%  of the expected precipitate mass for the five samples. The filter and precipitate was then laminated in a strip of packing tape. X-ray fluorescence was performed on the samples to confirm that the precipitate samples did indeed contain strontium. 

If $^{90}$Sr were present in the rainwater samples then the chemistry performed to produce a Sr precipitate would break the secular equilibrium between the decays of $^{90}$Sr (t$_{1/2}$ = 28.8 years) and $^{90}$Y (t$_{1/2}$ = 2.67 days).The samples were immediately counted after preparation and again one week later on a planar HPGe detector equipped with a thin beryllium entrance window (0.254 mm), which allows for transmission of low energy gamma and x-rays. The efficiency of a thin planar detector drops off rapidly for higher energy gammas ($>$100 keV). Betas however, maintain a constant efficiency at these higher energies since they are absorbed over a short distance within the HPGe crystal, so a $\beta$ particle above an MeV would have a far greater efficiency compared to gammas in that energy region. The dimensions of the planar HPGe used in these measurements was 32 mm in diameter and 13 mm thick. We then compared a ratio of the counts in the 0-0.546 MeV energy interval (the $^{90}$Sr endpoint) to that in the  0.546-2 MeV interval (approximate $^{90}$Y endpoint) immediately after filtering and again after one week to search relative changes between the two energy regions, which would be expected if $^{90}$Sr were present and feeding into $^{90}$Y decays. The immediate and one-week-old counting ratios are shown in {\bf Figure \ref{fig:ratiocomparison}}, which did not yield any indication for the presence $^{90}$Sr via $^{90}$Y growth.  The four samples were then stacked on top of each other for a week long counting to set limits, based upon $^{90}$Y $\beta$'s, calculated using Equation \ref{eq}, where N is the number of counts in the background, $\epsilon$ is the beta detection efficiency determined using a calibrated $^{90}$Sr source, V is the volume of rainwater (3.35 L) from which the precipitate was extracted, and LT is the livetime of counting. This produced a one sigma upper limit of 8.98 mBq/L for the presence of $^{90}$Sr in the rainwater samples analyzed.

\begin{equation}
{ A }_{ ul }\quad =\quad \frac { \sqrt { 2N }  }{ \epsilon \cdot V\cdot LT } 
\label{eq}
\end{equation}

\begin{table}[ht]
\centering
\scriptsize
\begin{tabular}{cccc} 
\hline
Isotope&Activity (Bq/L)&Fission Yield (\%) \citep{englandrider}&Volatility \citep{Sill198897,langowski} \\ \hline
$^{90}$Sr		& $<$ 8.99 $\times$ 10$^{-3}$ & 5.78 & Low\\
$^{137}$Cs	&0.136 $\pm$ 0.041 & 6.19		& High\\ \hline
\end{tabular}
\label{tab:volatility}
\caption{Volatility comparisons of $^{90}$Sr and $^{137}$Cs, alongside their approximate activities or limits set for their presences in rainwater from this work and  \citep{Norman2011}. These details are in support of $^{90}$Sr not being a major concern as a transported isotope to the Bay Area.}
\end{table}

\section{Soil and Sediment Samples}
Soil samples have been collected by LBNL personnel and analyzed at the LBF since the late 1950s.  The chosen analytic technique has been gamma ray spectroscopy, first using NaI(Tl) scintillation crystal detectors, but switching to high resolution HPGe semiconductor detectors in the 1980s when they became available with relative efficiencies of at least 30\%.  Initially, the motivation was to document the areal distribution in surface and near-surface soils of fission product radionuclides from atmospheric nuclear weapons testing.  Later applications included determination of the natural terrestrial radionuclides (U,Th,K), and determination of the radon emanation (principally the $^{238}$U series $^{222}$Rn) from these materials.  Analyses continue in all of the above applications.

The current study is designed to provide information for identifying and characterizing the radionuclides transported to the Berkeley Lab from the nuclear power plant disaster at Fukushima, Japan in the aftermath of the 3/11/2011 earthquake/tsunami. Two pairs of on-site soil samples provide some relevant information, as listed in {\bf Table \ref{tab:soil}}.  The first pair represents surface soils from a site near LBNL building 90 and a site near LBNL building 72, at which the surfaces had not been disturbed since before the onset of atmospheric nuclear weapons testing that began in the mid-1940s.  The second pair were sampled in late April 2011; one from the undisturbed site near LBNL building 90, and the other from the face of a shallow cut in deeply weathered Orinda Formation bedrock along one side of the parking lot at building 72, wherein the present surface has only been exposed since the late 1970s during a major expansion of the building.  

The absence of $^{134}$Cs in the 1998 samples identifies the activity of $^{137}$Cs remaining from mid 20$^{th}$ century atmospheric nuclear weapons tests. The results presented in the air monitoring results of this report indicate approximately equal activities of $^{134}$Cs and $^{137}$Cs were deposited at our sampling sites; hence the $^{134}$Cs activities observed in these April 2011 samples allows estimates to be made for the $^{137}$Cs activities delivered from the Fukushima accident.

A number of sediment samples obtained from laboratory roads were also analyzed. The asphalt road connecting building 72 with the main part of LBNL extends almost horizontally across a steep hillside.  An asphalt berm (curb) along the downslope edge of the road protects the slopes below from rainfall runoff from the road surface.  Partway along this stretch of roadway (near the turnoff to LBNL building 31) is a low spot at which particulates, mainly abraded from the road surface, collect against the curb.  Normal LBF procedure is to collect at least two samples per year from these deposits - - one just before the onset of the rainy season (September or October), and one at the end of the rainy season (May or June).  One to two kilograms of material is collected from the upstream end of one of these narrow deposits.  After air-drying, the material is sieved, and that fraction which passes through a 1/16Ó mesh screen becomes the sample to be analyzed.  Typically, the sieved material represents at least 80\% of the total collected material.  For analysis, these sediments are packed into one of the same counting containers as are used for soil samples.   These procedures were followed for the samples listed in {\bf Table \ref{tab:sediment}}.

\section{Food Measurements in October 2013}
The recent reports of radioactivity leaking from the Fukushima Nuclear Power Plant into 2013 prompted the measurement of various food samples to search for fission products, in particular fish since much of the media reports were related to continued releases of contaminated water into the ocean. There have already been studies of fission products in fish, such as those used to track migration in tuna using $^{134,137}$Cs by Madigan et. al. \citep{Madigan2012, Madigan2013}. 

To explore the presence of such products in the current food supply, fish from the Pacific Ocean and a few various other foods were obtained at local San Francisco Bay Area retail locations in September 2013. Samples containing a significant water weight, such as fish and yogurt, were first baked to reduce the overall sample mass and enable easier handling; but the wet weight was recorded for reporting and use in analysis. Samples were then placed in plastic bags and counted in a cylindrical geometry at the face of a HPGe detector (2.1 kg p-type HPGe crystal, 85\% relative efficiency) counted for a few days for each sample. For efficiency determinations, a calibration sample was created using a NIST Standard Reference Material-- `River Sediment 4350B,' which contains certified activities of $^{137}$Cs, $^{60}$Co, $^{152, 154}$Eu, $^{226}$Ra, $^{241}$Am, and others \citep{nist4350B}. The uncertainties of the isotopes in the river sediment vary from 4 to 21\%, where the primary isotope of interest for this set of measurements, $^{137}$Cs,  had an uncertainty of 6.4\%. To achieve a volume similar to the food samples analyzed, 53.5 grams of the river sediment was thoroughly mixed with 141.7 grams of flour and counted in the same geometry. 

\begin{figure}[!ht]
\centering
\includegraphics[width=\linewidth, angle=0]{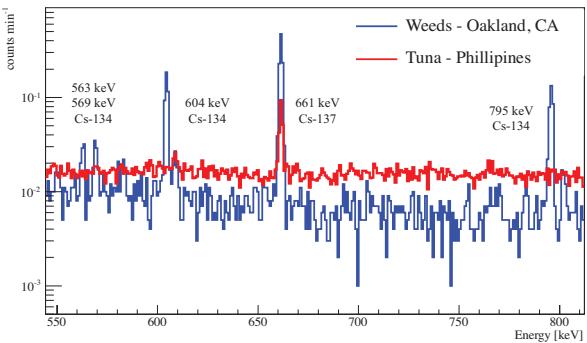}
\caption{A comparison of gamma spectra of local weeds (pictured in dotted blue) collected in April 2011 in Oakland, CA (recounted here in October 2013, at the LBNL LBF) along with a sample of Tuna from the Philippines (pictured in solid red) purchased locally at a San Francisco Bay Area retail location. $^{134}$Cs and $^{137}$Cs are readily visible and the remaining peaks are all natural background radioisotopes. The activities derived from these two samples are listed in {\bf Table \ref{tab:merweedstuna}} after recounting on this lower background system.}
\label{fig:weedstuna}
\end{figure}

The results of the food sampling are shown in {\bf Table \ref{tab:fukushimafood}}. In most of the Pacific Fish samples, $^{137}$Cs was present, with the highest activity found in tuna from the Philippines, which had a $^{137}$Cs activity of 0.24(4) Bq kg$^{-1}$. No samples had a detectable presence of $^{134}$Cs, which would have indicated that Fukushima products were present. Therefore, without $^{134}$Cs, the detected $^{137}$Cs is attributed to legacy activities such as surface nuclear weapons testing. It is also worth noting, that all samples had much higher levels $^{40}$K present, which is a naturally-occurring isotope. Comparing $^{137}$Cs to $^{40}$K is useful, since they both belong to the same column on the periodic table, and hence have similar affinities in various tissues and minerals. The tuna from the Phillipines for instance, had a $^{40}$K activity of 105(3) Bq kg$^{-1}$ -- more than four hundred times the activity of the $^{137}$Cs. Comparisons such as this are useful in evaluating the relative danger the $^{137}$Cs presents to the public without referencing limits set by regulations, as it allows one to make direct comparisons to natural background radiation. The last sample listed in {\bf Table \ref{tab:fukushimafood}} are weeds collected after the rainfall described by Norman in 2011 \citep{Norman2011} that had absorbed $^{134, 137}$Cs from Fukushima fallout. These weeds were cooked and washed very thoroughly in an attempt to determine whether the activity was inside the plants or merely surface-contaminated, and recounting proved that the activity was indeed absorbed inside the plant matter. The weeds were recounted in October 2013 along with the food samples in {\bf Table \ref{tab:fukushimafood}} as a comparison. Then the weeds were recounted again in November of 2013 along with the Philippines tuna sample as seen in {\bf Figure \ref{fig:weedstuna}}, this time on a lower background system; the same 115\% HPGe system at the LBNL LBF used in the air monitoring efforts. The results of the low background counting of these samples are seen in {\bf Table \ref{tab:merweedstuna}}. These weeds were used as a proxy for Fukushima fallout-- the ratio of $^{134}$Cs/$^{137}$Cs was compared to the $^{137}$Cs we detected in our food samples, which confirmed that if the $^{137}$Cs was indeed from Fukushima, we should have easily seen $^{134}$Cs over our detection limits. Therefore, the only detected $^{137}$Cs was from pre-Fukushima, legacy sources. The U.S. Food and Drug Administration (FDA) Derived Intervention Level (DIL) for the combined $^{134,137}$Cs activity in food \citep{fdadil} is currently 1200 Bq kg $^{-1}$. All food samples measured were more than 1000-times smaller than the 1200 Bq kg $^{-1}$ FDA DIL and pose no concern to the public and were far less than natural gamma-emitting radioisotopes present, namely $^{40}$K. Other studies that have found Fukushima isotopes in fish, such as Madigan, et. al.\citep{Madigan2012, Madigan2013}, have gone a step further and to show that Fukushima-sourced dose rates due to ingestion of even their highest-activities of $^{134,137}$Cs found in tuna are absolutely minuscule in comparison to the natural dose from $^{210}$Po as discussed in great detail in Fisher et. al. \cite{Fisher03062013}.

\begin{table}[ht]
\centering
\begin{tabular}{cccc} 
\hline
Sample				&	$^{134}$Cs			&	$^{137}$Cs		& $^{40}$K	\\
					& Bq kg$^{-1}$  &	Bq kg$^{-1}$			&	Bq kg$^{-1}$			\\ \hline
Local Weeds 04/2011	&	7.0(5)		&	20.8(7) &  294(8)	\\
Tuna Phillipines	&	 $<$0.07			&	 0.20(1) &   95.3(6) 	\\ \hline
\end{tabular} 
\caption{The results of recounting the local weeds and the tuna sample from the Phillipines on a lower background system at the LBF, the spectra of which are seen in {\bf Figure \ref{fig:weedstuna}}. Since the weeds were collected in April of 2011 after absorbing Fukushima fallout, the ratio of $^{134}$Cs to $^{137}$Cs was used as a proxy to compare to the tuna sample, which had a consistent limit for $^{134}$Cs  detection as compared to the $^{137}$Cs present. This suggests that if the $^{137}$Cs in the tuna sample was indeed from Fukushima, then our system would have seen $^{134}$Cs as it would be present above the listed detection limits. Therefore the $^{137}$Cs is from legacy, pre-Fukushima sources.}
\label{tab:merweedstuna}
\end{table}

\section{Chernobyl Comparisons}

\subsection{Chernobyl Air Monitoring}
\begin{figure}[!ht]
\centering
\includegraphics[width=\linewidth, angle=0]{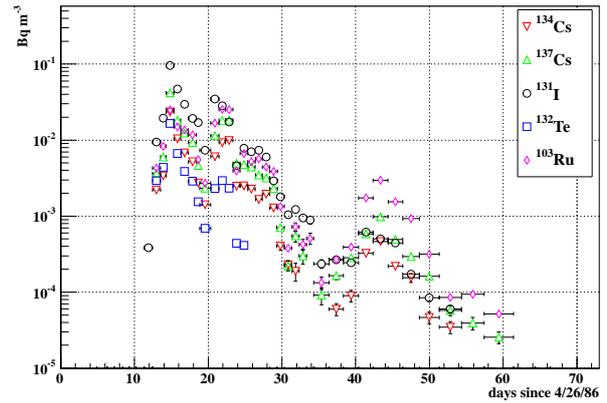}
\caption{Airborne concentrations fission products starting April 26, 1986 following the Chernobyl disaster as measured on a HEPA filter at the LBF in Berkeley, CA. Horizontal error bars, when visible, represent filter exposure periods. Missing error bars are smaller than the visible data marker. Plotted uncertainties seen here are statistical only, and a conservative systematic uncertainty of $\sim$10\% should also be assumed.}
\label{fig:chernobylfallout}
\end{figure}

Similar air monitoring efforts were also made by the Low Background Facility in the aftermath of the Chernobyl accident in 1986. Following the accident an air filtration system was installed and operated. This system had a variable motor speed that was manually set to draw 2 CFM of air through filters 5 inches by 9 inches in size. The filters were then folded in half and counting in the annulus of a Marinelli style beaker, such that they were wrapped flat along the side of a HPGe spectrometer (p-type, 30\% relative efficiency). Filters were typically exposed in 24 hour intervals and counted immediately after removal. Results of the airborne monitoring performed in the aftermath of the Chernobyl incident and are displayed in {\bf Figure \ref{fig:chernobylfallout}}. In addition to the isotopes seen from Fukushima, there was also an extended release of $^{103}$Ru (t$_{1/2}$ = 39.26 days, primary gamma ray of 497.9 keV at 90.9\% intensity). The results of this monitoring showed the highest activities for fallout in the Bay Area on May 5, 1986  with airborne activities of:  $^{132}$Te - 16.6$\pm$0.1 mBq m$^{-3}$, $^{131}$I - 95.6$\pm$0.5 mBq m$^{-3}$, $^{134}$Cs - 23.4$\pm$0.3 mBq m$^{-3}$, $^{137}$Cs - 41.8$\pm$0.4 mBq m$^{-3}$, and $^{103}$Ru - 25.3$\pm$0.5 mBq m$^{-3}$ (statistical uncertainties only, a $\sim$10\% systematic uncertain should also be assumed).

When comparing the highest levels of fallout radionuclides from the Chernobyl and Fukushima accidents we can easily see the Chernobyl activities measured at the San Francisco Bay Area of California were about an order of magnitude higher than those from Fukushima. Roughly the same seven-day temporal period appears to separate successive maxima for these fallout nuclides, seen most clearly in {\bf Figure \ref{fig:chernobylfallout}}, a graphical presentation of the Chernobyl data, and {\bf Figure \ref{fig:allfalloutfull}} for those from Fukushima.

\subsection{Chernobyl Food Monitoring}
Also measured in the late 1980's were a series of imported and domestic food products. The study from Norman and Harvey \citep{Norman88, Norman89} was previously unpublished and is summarized here. After hearing extended reports of fallout isotopes originating from the Chernobyl disaster appearing in foods, a small survey of approximately 50 imported foods available in the San Francisco Bay Area was conducted. The samples originated from 21 European countries as well as 4 from the U.S.A. The study was conducted with a 109 cm$^{3}$ (approximately 0.6 kg) HPGe detector. All samples were counted directly on the face of the detector and all had masses between 10-500 grams. Efficiencies and attenuation affects were measured using calibrated point sources in various positions with varying thicknesses of absorbers to determine a counting efficiency to $\pm$20\%.  A table of results is presented in {\bf Table \ref{tab:chernobylfood}}.

Many of the products showed no activity above background and of those that did show radioisotopes,  $^{134}$Cs and $^{137}$Cs were the only residual fission products still present from Chernobyl as others had decayed away. In order to confirm the origin of the two isotopes, a ratio of the $^{137}$Cs/$^{134}$Cs activities were used to estimate the age they were produced. All food product samples, except for one, produced a $^{137}$Cs/$^{134}$Cs ratio of  2.95$\pm$0.30 (as of October 1987). Due to the difference in half-lives of the isotopes, this ratio can be used as an identifier of their source, since it will be unique for each reactor in its fuel cycle. Decay correcting this ratio to the date of the Chernobyl incident, April 25, 1986, the ratio is 1.88$\pm$0.19 which was in agreement with ratios present in prompt fallout such as those in {\bf Figure \ref{fig:chernobylfallout}}, which yield a ratio of 1.90$\pm$0.08. All food measurements were below the maximum permissible limits allowed in food at the time, which was 10 pCi g$^{-1}$ (370 Bq kg$^{-1}$) after being reduced from 75 pCi g$^{-1}$ (2.8 $\times 10^{3}$ Bq kg$^{-1}$) \citep{usda}. The authors extrapolated these measurements to exposure, if one were to eat 2 kg/day of foods at this level of activity. Factoring in that both of the isotopes have a biological half-life of 70 days \citep{lanl}, they estimated that at equilibrium one would have a total activity of 0.5 $\mu$Ci (2 $\times 10^{7}$ Bq kg$^{-1}$) of $^{134}$Cs and 1.5 $\mu$Ci (5.6 $\times 10^{7}$ Bq kg$^{-1}$) of $^{137}$Cs, which were well below the limits of 2.0 $\mu$Ci  (7.4 $\times 10^{7}$ Bq kg$^{-1}$) of $^{134}$Cs and 3.0 $\mu$Ci  (1.1 $\times 10^{8}$ Bq kg$^{-1}$) of $^{137}$Cs set for non-radiation workers at the time \citep{lanl}.

\section{Conclusion}
Monitoring of fallout from the Fukushima nuclear accident was performed on a variety of media at the LBNL Low Background Facility starting soon after the tsunami on March 14, 2011 and reported here through the end of 2012 on air filters, automobile air filters, and rainwater. At the local LBNL LBF location, HEPA air filters were used in sampling durations ranging from 24 hours to one month while monitoring radioisotope concentrations. Rainwater was also collected and analyzed through the end of 2012 in Oroville, CA. Additionally, the fission products monitored provided a useful demonstration of the use of automobile filters as a low cost means of monitoring for radioisotopes. More extensive analysis was performed upon rainwater samples from March 2011 that contained the highest measured activities in the Berkeley area in an effort to search for the presence of $^{90}$Sr in the rainwater that arrived in the initial weeks following the incident, in which it was not detected and one sigma limits were placed at $<$ 8.98 mBq L$^{-1}$. A series of food measurements were also performed in October 2013 upon imported food products from the Pacific region, and although background $^{137}$Cs was present from sources prior to Fukushima, it was far below any level of concerns due to radioactivity-- both from FDA DIL's and comparison to natural $^{40}$K. Measurements made by the same authors of air filters and food products in the aftermath of the Chernobyl disaster are also presented here in comparison to the Fukushima fallout. The main conclusion drawn from these sets of data is that the peak fallout activities from Chernobyl in 1986 upon the San Francisco Bay Area were approximately an order of magnitude more than the levels seen from Fukushima in 2011. 

The fission products measured in this work in the San Francisco Bay Area of California were not found at any time to be of concern to public or environmental health. These conclusions are justified not only through governmental limits of exposure and activities, but also by direct comparison of natural radioactivities also present in all samples measured. 

\section{Acknowledgements}
We wish to thank M. B. Norman for useful discussions that prompted our searches for Fukushima fallout in food samples.

This material is based upon work supported by the Department of Energy National Nuclear Security Administration under Award Number(s) DE-NA0000979 and by the Director, Office of Energy Research, Office of High Energy and Nuclear Physics, Division of Nuclear Physics, of the US Department of Energy under Contract No. DE-AC02-05CH11231. 

\section{References}
\bibliographystyle{model1a-num-names}
\bibliography{bibtex}

\appendix
\section{Supplementary Figures and Tables}

\begin{table}[ht]
\centering 
\scriptsize
\begin{tabular}{rclc}
\hline
Isotope	&	Peak keV (intensity)	&	Source	&	t$_{1/2}$	\\ \hline
$^{132}$Te	&	228 (88\%)	&	fission	&	3.2 days	\\
$^{212}$Pb	&	238 (43\%)	&Th series	&	10.64 hours	\\
$^{131}$I	&	284 (6\%), 364 (82\%),	 	&	fission	&	8.02 days	\\
		&	 637 (7\%), 723 (2\%)	 	&&		\\
$^{214}$Pb	&	352 (38\%)	&	U series	&	26.8 min	\\
$^{136}$Cs	&	340 (42\%), 819 (100\%)	&	fission	&	13.16 days	\\
$^{7}$Be	&	477 (11\%)	&	cosmogenic	&	53.12 days	\\
n/a	&	511	&	annihilation	&	n/a	\\
$^{132}$I	&	522 (16\%), 667 (97\%), 	&	fission	&	2.295 hours	\\
	&773 (76\%)	&		&		\\
$^{134}$Cs	&	563 (8\%), 569 (15\%), 	&	fission	&	2.06 years	\\
	&	 604 (98\%), 795 (86\%)	&		&	\\
$^{208}$Tl	&	583 (84\%)	&	Th series	&	3.05 min	\\
$^{214}$Bi	&	609 (46\%)	&	U series	&	19.9 min	\\
$^{137}$Cs	&	661 (85\%)	&	fission	&	30.07 years	\\ 	
\hline
\end{tabular}	
\caption{Radioisotopes shown in Figures \ref{fig:airfilter} and \ref{fig:autofilter}, along with the gamma ray energies, intensities, and half-life. Isotope data adapted from \citep{toi}.}
\label{tab:isotopes}
\end{table}

\begin{figure}[!ht]
\centering
\includegraphics[width=\linewidth, angle=0]{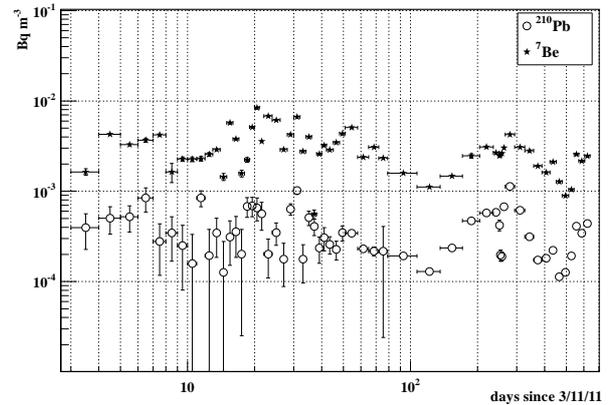}
\caption{Airborne concentrations of two naturally-produced radioisotopes, $^{210}$Pb and $^{7}$Be, from  March 11, 2011 to the end of 2012 as measured on HEPA filters at the LBF in Berkeley, CA. These isotopes provide some context to the scale of the activities seen from the fission products in Figure \ref{fig:allfalloutfull}. Horizontal error bars, when visible, represent filter exposure periods. Missing error bars are smaller than the visible data marker.Plotted uncertainties seen here are statistical only, and a conservative systematic uncertainty of $\sim$10\% should also be assumed.}
\label{fig:pbbe}
\end{figure}

\begin{figure}[!ht]
\centering
\includegraphics[width=\linewidth, angle=0]{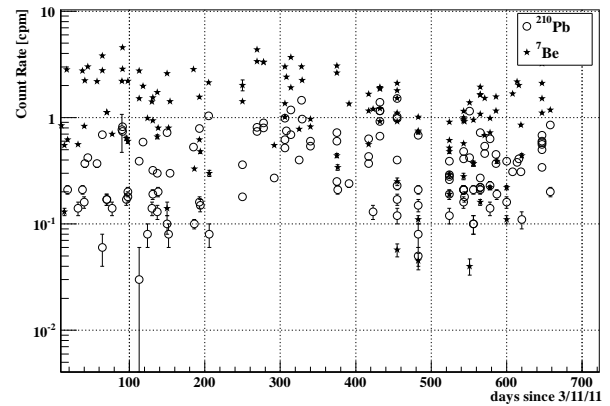}
\caption{Counting rates for natural radioisotopes as seen on automobile filters in Berkeley, CA. These gross counting rates are of use for comparison to the counting rates in Figure \ref{fig:auto}, which reveals the fallout counting rates were similar in scale to natural radioisotopes. The day plotted for filters represents the date after 3/11/11 when the filter was removed from a vehicle. Counting rates were not converted to airborne activity, as the full mileage (and hence the airflow) information was not available for the full data set, so the information is presented solely as a qualitative demonstration of the use of automobile filters for radionuclide monitoring.}
\label{fig:autonatural}
\end{figure}

%\begin{figure}[!ht]
%\centering
%\includegraphics[width=\linewidth, angle=0]{OROrainAll-loglin-first20days}
%\caption{Observation of Fukushima fallout in rainwater collected in Oroville, CA during the initial arrival. $^{7}$Be is shown as a comparison to a naturally-produced airborne isotope that behaves as `fallout.' Horizontal error bars, when visible, represent filter exposure periods. Missing error bars are smaller than the visible data marker. Plotted uncertainties seen here are statistical only, and a conservative systematic uncertainty of $\sim$10\% should also be assumed.}
%\label{fig:ororain20}
%\end{figure}

\begin{figure}[!ht]
\centering
\includegraphics[width=\linewidth, angle=0]{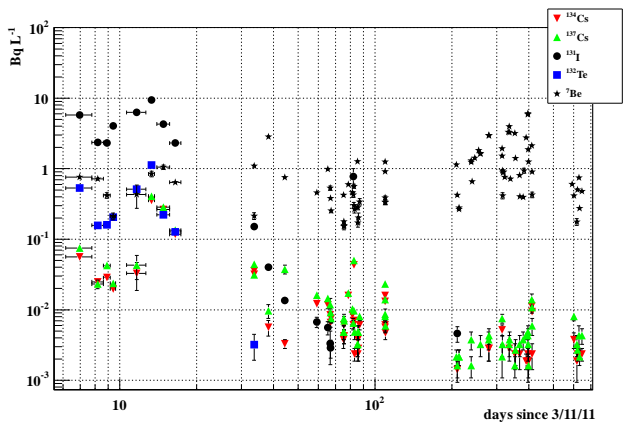}
\caption{Observation of Fukushima fallout in rainwater collected in Oroville, CA over the entire monitoring duration to the end of 2012. $^{7}$Be is shown as a comparison to a naturally produced airborne isotope. Horizontal error bars, when visible, represent filter exposure periods. Missing error bars are smaller than the visible data marker. Plotted uncertainties seen here are statistical only, and a conservative systematic uncertainty of $\sim$10\% should also be assumed.}
\label{fig:ororainloglog}
\end{figure}

\begin{figure}[!ht]
\centering
\includegraphics[width=0.7\linewidth, angle=0]{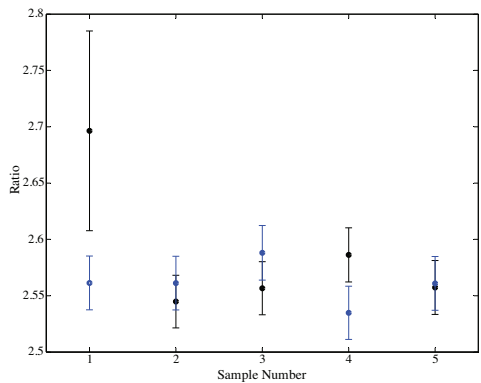}
\caption{Low energy to high energy ratios (0 to 546 keV region to 546 to 2000 keV region) for Sr precipitate samples immediately after preparation (black) and at one week old (blue). This showed no evidence or the presence of $^{90}$Sr since all measurements agree to within approximately 1$\sigma$.}
\label{fig:ratiocomparison}
\end{figure}

\begin{table}[ht]
\centering
\begin{tabular}{cccc} 
\hline
 Location & Date	&	$^{134}$Cs &	$^{137}$Cs 			\\
	& m/d/yr  &	Bq kg$^{-1}$			&	Bq kg$^{-1}$			\\ \hline
bldg 72 vicinity&7/25/98	&	$<$0.01			&	1.06(5)	\\
bldg 90 vicinity&8/5/98	&	$<$0.01			&	8.5(1)	\\
bldg 72 vicinity&4/27/11	&	1.70(4)	&	2.69(5)	\\
bldg 90 vicinity&4/30/11	&	0.28(2)	&	3.47(5)	\\ \hline
\end{tabular} 
\caption{Sample results of surface soil tests at LBNL in 1998 and post-Fukushima in spring 2011.}
\label{tab:soil}
\end{table}

\begin{table}[ht]
\centering
\scriptsize
\begin{tabular}{cccccc} 
\hline
 Date		&	$^{134}$Cs	&	$^{137}$Cs 	&	$^{131}$I 	&	$^{7}$Be 	&	$^{210}$Pb 	\\
 m/d/yr	&	Bq kg$^{-1}$	&	Bq kg$^{-1}$	&	Bq kg$^{-1}$	&	Bq kg$^{-1}$	&	Bq kg$^{-1}$	\\ \hline
11/18/10	&	$<$0.02	&	0.09(1)	&	$<$0.01	&	84.8(4)	&	53.5(6)	\\
3/1/11		&	$<$0.02	&	0.11(1)	&	$<$0.02	&	220.0(8)	&	46.6(7)	\\
4/11/11		&	20.0(1)	&	24.6(1)	&	21.29(8)	&	318.9(8)	&	59.0(9)	\\
6/9/11		&	58.6(2)	&	73.8(2)	&	0.15(4)	&	346.7(8)	&	92(2)	\\
9/27/11		&	34.5(1)	&	48.2(1)	&	$<$0.03	&	88.8(4)	&	71.9(9)	\\
10/9/11		&	22.6(1)	&	31.8(1)	&	$<$0.03	&	90.8(4)	&	52.5(9)	\\
1/4/12		&	17.43(5)	&	26.5(1)	&	$<$0.02	&	67.8(4)	&	48.8(6)	\\
4/9/12		&	15.67(5)	&	26.1(1)	&	$<$0.02	&	163.1(4)	&	61.6(7)	\\
9/20/12		&	14.44(5)	&	27.3(1)	&	$<$0.02	&	51.4(3)	&	82.8(6)	\\
11/25/12		&	5.26(5)	&	10.6(1)	&	$<$0.02	&	104.5(4)	&	43.3(6)	\\ \hline
\end{tabular} 
\caption{Sample results of surface sediment tests along the roadside at LBNL building 72 post-Fukushima in spring 2011. The activities in this sediment often become elevated since rainwater drains off the roadway and a natural filtration mechanism occurs.}
\label{tab:sediment}
\end{table}

\begin{table}[ht]
\centering
\scriptsize
\begin{tabular}{llccc}
\hline
Sample	&	mass 	&	$^{40}$K 	&	$^{134}$Cs 	&	$^{137}$Cs	\\  
		&	g	&	Bq kg$^{-1}$	&	Bq kg$^{-1}$&	Bq kg$^{-1}$	\\\hline
Tuna Philippines	&	444.5	&	105(4)	&	n.d.	&	0.27(6)	\\
Fiji Tuna	&	506	&	62(2)	&	n.d.	&	0.12(2)		\\
Local Red Snapper	&	440	&	67(3)	&	n.d.	&	0.10(3)		\\
Local Organic Yogurt	&	907.2	&	66(2)	&	n.d.	&	n.d.		\\
Hawaii Swordfish	&	467.2	&	99(4)	&	n.d.	&	n.d.		\\
Local Squid	&	902.6	&	21(1)	&	n.d.	&	n.d.		\\
Alaskan Salmon	&	430.9	&	82(3)	&	n.d.	&	n.d.		\\
Alaskan Cod	&	444.5	&	68(3)	&	n.d.	&	0.13(4)		\\
Atlantic Mackerel	&	396	&	78(3)	&	n.d.	&	n.d.		\\
Japanese Dried Bonito	&	100	&	195(10)	&	n.d.	&	n.d.		\\
Local Grape Leaves	&	64.9	&	249(14)	&	n.d.	&	n.d.		\\
Local Weeds (04/2011)	&	30.6	&	294(27)	&	7.0(6)	&	18(1)		\\ \hline
\end{tabular} 
\caption{Food measurements made in search of fallout from the Fukushima nuclear accident in food samples purchased in San Francisco Bay Area retail locations in October 2013, with the exception of the local weeds that were extracted during April 2011, after the initial rainfall containing fallout isotopes, but re-counted here in October 2013. Activities of $^{137}$Cs are listed alongside natural $^{40}$K activity that were also measured. Reported masses for the food samples are the wet weight before baking to expel excess water mass. For entries with no data, it can be assumed the value is less than the approximate MDA's for this set of measurements, which varied over the following ranges: 0.06-0.15 Bq kg$^{-1}$ for $^{137}$Cs and 0.05-0.13 Bq kg$^{-1}$ for $^{134}$Cs. The absence of $^{134}$Cs suggests that the $^{137}$Cs measured in some of the samples is from legacy sources prior to Fukushima. }
\label{tab:fukushimafood}
\end{table}

\begin{table}[ht]
\centering
\scriptsize
\begin{tabular}{llcc}
 \hline
Country of Origin & Product & \multicolumn{2}{c}{$^{134,137}$Cs combined activity}\\
	&		&	pCi g$^{-1}$	&	Bq kg$^{-1}$	\\  \hline
England	&	beer	&	n.d.	&	n.d.	\\
	&	crackers	&	n.d.	&	n.d.	\\
	& 			&		&		\\
Belguim	&	beer	&	n.d.	&	n.d.	\\
	&	shallots	&	n.d.	&	n.d.	\\
	& 			&		&		\\
Spain	&	white wine (1986)	&	n.d.	&	n.d.	\\
	& 			&		&		\\
France	&	red wine	&	n.d.	&	n.d.	\\
	&	white whine	&	n.d.	&	n.d.	\\
	&	black olives	&	n.d.	&	n.d.	\\
	&	raspberry jam	&	n.d.	&	n.d.	\\
	&	apricot jam	&	n.d.	&	n.d.	\\
	&	honey (1986)	&	0.4	&	14.8	\\
	&	goat cheese	&	0.2	&	7.4	\\
	&	rouquefort cheese	&	0.2	&	7.4	\\
	& 			&		&		\\
Italy	&	pasta (brand 1)	&	n.d.	&	n.d.	\\
	&	pasta (brand 2)	&	1.3	&	48.1	\\
	&	pasta (brand 3)	&	0.8	&	29.6	\\
	&	pasta (brand 4)	&	1.6	&	59.2	\\
	&	olive oil	&	n.d.	&	n.d.	\\
	&	red wine (1986)	&	n.d.	&	n.d.	\\
	&	white wine (1986)	&	n.d.	&	n.d.	\\
	&	dried mushrooms	&	4	&	148	\\
	& 			&		&		\\
Switzerland	&	green cheese w/herbs	&	0.2	&	7.4	\\
	&	emmenthaler cheese	&	0.13	&	4.81	\\
	& 			&		&		\\
Austria	&	blackberry jam	&	0.4	&	14.8	\\
	&	beer	&	n.d.	&	n.d.	\\
	& 			&		&		\\
West Germany	&	cheese w/herbs	&	n.d.	&		\\
	&	soft cheese	&	n.d.	&	n.d.	\\
	&	beer	&	n.d.	&	n.d.	\\
	&	gummi bear candy	&	n.d.	&	n.d.	\\
	& 			&		&		\\
The Netherlands	&	beer	&	n.d.	&		\\
	& 			&		&		\\
Norway	&	beer	&	n.d.	&	n.d.	\\
	&	goat cheese	&	5	&	185	\\
	& 			&		&		\\
Sweden	&	lingon berry sauce	&	0.13	&	4.81	\\
	&	crispbread	&	n.d.	&	n.d.	\\
	& 			&		&		\\
Finland	&	rye crackers	&	0.16	&	5.92	\\
	& 			&		&		\\
U.S.S.R.	&	vodka	&	n.d.	&	n.d.	\\
	& 			&		&		\\
Czechoslovakia	&	beer	&	n.d.	&	n.d.	\\
	& 			&		&		\\
Poland	&	ham	&	n.d.	&	n.d.	\\
	& 			&		&		\\
Hungary	&	paprika	&	n.d.	&	n.d.	\\
	& 			&		&		\\
Roumania	&	feta cheese	&	n.d.	&	n.d.	\\
	& 			&		&		\\
Yugoslavia	&	beer	&	0.1	&	3.7	\\
	& 			&		&		\\
Greece	&	black olives	&	n.d.	&	n.d.	\\
	&	beer	&	1.4	&	51.8	\\
	& 			&		&		\\
U.S.A.	&	beer	&	n.d.	&	n.d.	\\
	&	pasta	&	n.d.	&	n.d.	\\
	&	cream cheese	&	n.d.	&	n.d.	\\
	&	cheddar cheese	&	n.d.	&	n.d.	\\
\end{tabular}
\caption{Chernobyl food measurements from 1988 previously unpublished (formally) from Norman and Harvey \cite{Norman88, Norman89}. All samples were purchased at retail locations in the San Francisco Bay Area. Overall uncertainties are estimated to be $\pm$25\%. n.d. indicates no activity was detected, in which the detection limits for this study varied from 0.02 - 0.1 pCi/g (0.74 - 3.7 Bq/kg) amongst the samples.}
\label{tab:chernobylfood}
\end{table}

%\appendix
%\include{supplementary}
%\include{airfiltertablecopy}  %removed to save space
%\include{ororaintablecopy}  %removed to save space
%\include{chernobyltable}  %removed for saving space

%%%%%%%%%%%%%%%%%%%%%%%%%%%%%%%%%%%%%%%%%%%%%%%%%%%%%%%%%%%%%%%%%

%% References
%%   \citep{key}          ==>>  [#]






% \end{thebibliography}

\end{document}